\documentstyle[12pt,epsf,amssymb]{article}
\begin{document}
\title{\normalsize{\bf COVARIANCE, CORRELATION AND ENTANGLEMENT} }
\author{{\small R.I.A. DAVIS, R. DELBOURGO and P.D. JARVIS} \\
{\small School of Mathematics and Physics}\\
{\small University of Tasmania, Hobart, Australia}}

\def\half{\frac{1}{2}}

\date{}
\maketitle
\begin{abstract}
Some new identities for quantum variance and covariance involving
commutators are presented, in which the density matrix and the operators
are treated symmetrically. A measure of entanglement is proposed for
bipartite systems, based upon covariance. This works for two- and
three- component systems but produces ambiguities for multicomponent systems
of composite diemnsion. Its relationship to angular momentum dispersion for 
symmetric spin states is described.
\end{abstract}

\section{Introduction}
Several measures of entanglement \cite{entanglement} or quantum correlations 
have been proposed: some are associated with the preparation of the state, 
others with the process of purification or distillation \cite{purification} 
and yet others with the notion of mutual information or relative entropy
\cite{relentropy}. In this paper we wish to suggest another measure, based
on covariance, in which the acts of state creation and observation are 
considered in a dual manner.

In practice it is very natural to describe the condition of the system (its
method of preparation or lack of it) in terms of a density matrix which is 
tied to the subsequent observations on it. This is how the linkage between 
observer and observed occurs quantum mechanically, and of course the results 
are expressed in terms of traces over appropriate functions of the density 
matrix and of the operators being measured \cite{corr}. Indeed Mermin 
\cite{Mermin} has taken that view that the density matrix, and the 
correlations between observables which thereby ensue, constitute {\em all} of 
physical reality.

In this paper we will also focus on the density matrix. Because binning of
observations is a necessity in practice, the dimension of the density matrix 
is thereby determined: $N$ separate bins produce a density matrix $\rho$ that
is an $N\times N$ hermitian matrix, satisfying the usual hermiticity and trace
conditions. In this way, we can regard the basis as an $N$-level system, 
rather like a particle of angular momentum $J = (N-1)/2$. Hence, although we
might be studying the probability distribution of an observable which 
actually possesses a continuous spectrum, we can still regard it as a 
spin-like system; in practical terms, the bigger the binning number $N$, 
the greater the precision of the information about the continuous variable,
but obviously $N$ is {\em never infinite}. For spin measurements, we need not 
go to such pains because $N$ is fixed for us at the start.

With the focus on density matrices, we will carry out measurements (without
mutual interference) on two subsystems, 1 and 2 say, so their corresponding 
observables, superscripted by (1) and (2), are commuting operators. The system
will be separable \cite{separability} or ``disentangled'' if the larger 
density matrix $\rho$ is merely a direct product of density matrices 
associated with the two subsystems, or $\rho=\rho^{(1)}\otimes\rho^{(2)}$;
a particular case arises when the initial or prepared state is the direct
product of two subsystem states, $|\psi\rangle =|\phi^{(1)}\rangle |\chi^{(2)}
\rangle$. When two subsystems are disentangled, the results of measuring
any quantity $A^{(1)}$ in the first subsystem are not tied to the results of
measuring any quantity $B^{(2)}$ in the second subsystem; necessarily
$$\langle A^{(1)}B^{(2)}\rangle=\langle A^{(1)}\rangle\langle B^{(2)}
\rangle,$$
for all choices of $A$ and $B$. However, if $\rho\neq\rho^{(1)} \otimes 
\rho^{(2)}$, the configuration is non-factorizable and the covariance, 
\begin{equation}
{\rm cov}(A^{(1)}B^{(2)})\equiv \langle A^{(1)}B^{2)}\rangle -
\langle A^{(1)}\rangle\langle B^{(2)}\rangle,
\end{equation}
no longer disappears.

The real issue is how to quantify the entanglement or lack of factorizability 
\cite{separability} of the larger density matrix. Several proposals have been 
advanced in the literature, but none of them is entirely simple or definitive
\cite{entanglement}. However all researchers in this field seem to agree 
on the following three conditions for an entanglement measure $E(\rho)$:
\begin{enumerate}
\item $E(\rho)=0$ iff $\rho$ is separable, ie if the density matrix can be
written as $\rho = \sum_i p_i\rho^{(1)}_i\otimes\rho^{(2)}_i$.
\item Local unitary transformations should leave $E(\rho)$ invariant.
\item $E(\rho)$ should not increase under local measurement and classical
communication procedures, we intuitively know that such procedures 
cannot add non-locality characteristics to the system being measured.
\end{enumerate}
As an extra requirement, it would be nice if $E(\rho)$ gave some indication 
of the extent of violation of Bell-type inequalities \cite{corr}.

In this paper we want to put forward a concrete scheme for quantifying 
correlations between two subsystems and their possible entanglement. 
The scheme is based on a generalization of eq. (1) and particular 
choices of operators $A^{(1)}$ and $B^{(2)}$, which are readily applicable
and rooted in the density matrix notion. In the next section we discuss 
several matters connected with non-separability of states and their
influence on subsequent subsystem measurements. Because we deal with
practical observations, the density matrix is truly discrete and we can assume
that the elements of the vector space on which it lives have equal weight. 
As already mentioned, one may regard the dimension $N = N^{(1)}+N^{(2)}$ as
corresponding to a ``spin system'', with each component {\em carrying equal 
weight}, and can adopt the same stance for the subsystem dimensions 
$N^{(1,2)}$. (This restriction can be relaxed if the components have unequal 
weights, such as atomic energy levels at a {\em finite} temperature.)

The next section deals with the generalities of simultaneous measurements
and their covariance properties \cite{covariance}. This is followed by our 
suggestion for quantifying entanglement of two subsystems within a larger 
entity, which is shown to be consistent with normal expectations for two spin 
1/2 subsystems, when $N^{(1)}=N^{(2)}=2$. We also discuss the use of total 
spin dispersion \cite{DeltaJ} as another measure of entanglement, with an 
allied appendix concerning the Majorana-Penrose \cite{MP} representation of 
spin states on the Poincar\'{e} sphere. The subsequent sections deal with 
entanglement measures for larger 
value of $N^{(1)}$ and $N^{(2)}$. Finally we discuss general questions 
pertaining to our suggested measure; these include Rovelli's notion that 
information in quantum mechanics is relational \cite{relentropy}, Mermin's 
notions of correlations between local observables \cite{Mermin}, and the 
difference between our modified correlation measure with classical 
correlations for impure states.

\section{Correlations and density matrices}
Elementary texts on quantum mechanics teach us that the results of all
physical measurements and processes can be tied to the evaluation of traces
of products of observables with the hermitian density matrix $\rho$. Thus
statistical formulae like
$$\langle F \rangle = {\rm Tr}(\rho F);\quad {\rm Tr}(\rho^2)\leq 1, $$
etc.\, are part of the standard repertoire. Of course, the $\rho$-eigenvalues
lie between 0 and 1; in the latter case we are dealing with a pure state
when the density matrix reduces to a projector $\rho\rightarrow P_\psi\equiv|
\psi\rangle\langle\psi|$, while the most random situation $\rho = 1/N$
corresponds to the case of maximum entropy.

The covariance for any two commuting observables $A,B$ in a mixed state $\rho$
is defined as
\begin{equation} 
 {\rm cov}_{\rho}(A,B)\equiv\langle A B\rangle-\langle A\rangle
 \langle B\rangle={\rm tr}(\rho AB)-
 {\rm tr}(\rho A){\rm tr}(\rho B).
\end{equation}
Clearly, ${\rm var}_{\rho}(A)={\rm cov}_{\rho}(A,A)$. Less well-known is the 
fact that pure state dispersions and correlations can be neatly expressed in 
terms of a single trace. Consider the quantity
\begin{equation}
 C_{\rho}(A,B)\equiv {\rm tr}([\rho,A][B,\rho])/2 =
 {\rm tr}(\rho^2\{A,B\}/2 - \rho A\rho B),
\end{equation}
where $A$ and $B$ are {\em any} two operators. This quantity will be referred 
to as the {\em alternative covariance}
\footnote{Evaluating traces of larger numbers of pure state commutators, one 
may establish algebraically that for odd numbers of products, the traces do
vanish. For instance, ${\rm tr}\left([A,\rho][B,\rho][C,\rho]\right)=0$, etc.}.

We now present some elementary results about $C_\rho$ which follow simply from
this definition:
\begin{enumerate}
\item $C_\rho(A,B) = C_\rho(A-a,B-b)$, where $a,b$ are any two constants.
\item $C_\rho(aA,bB) = abC_\rho(A,B)$.
\item $C_\rho(\sum_i A_i,\sum_j B_j) = \sum_{i,j} C_\rho(A_i.B_j).$
\item $C_\rho(A,A) = {\rm tr}(\rho^2 A^2 -\rho A\rho A).$
\item $C_{U\rho U^{\dag}}(A,B)=C_\rho(U^{\dag}AU,U^{\dag}BU)$, where $U$ is
      any unitary transformation. Thus a change of basis for the state is
      equivalent to an inverse change of basis for the operators.
\item $C_\rho(A,A^{\dag})C_\rho(B,B^{\dag})\geq |C_\rho(A,B^{\dag})|^2$. This
      follows by considering the operator $T= [\rho,A - cB]$, with
      $c={\rm tr}([\rho,A][A^{\dag},\rho])/{\rm tr}([\rho,A][B^{\dag},\rho],$
      and noting that ${\rm tr}(TT^{\dag}) \geq 0.$
\item $C_\rho(A,A^{\dag})$ is {\em real}.
\item $|C_\rho(A,B)|^2$ is symmetrical under interchange, conjugation and
      change of phase of the two operators.
\end{enumerate}
All of these properties are shared by the usual covariance cov$_\rho(A,B)$.
Nevertheless, alternative covariance $C_\rho$ does not provide an indication 
of variance and covariance in the usual sense. For instance, if the state is 
one of maximum entropy, on the one hand we have $C_\rho(A,B) = 0$ for all 
$A, B$ because $\rho$ is proportional to unity; on the other hand, the
${\rm cov}_\rho(A,B)$ need not vanish.

\noindent Some special cases for the operators $A, B$ can now be studied.  
\begin{enumerate}
\item If $A$ and $B$ commute, $C_\rho(A,B) = {\rm tr}(\rho^2AB-\rho A\rho B).$
\item If $A$ and $B$ are both hermitian, $C_\rho(A,B)$ becomes real.
\item If $A$ and $B$ are both unitary, $C_\rho(A,A^{\dag})={\rm tr}(\rho^2 -
      \rho A\rho A^{\dag}) \leq {\rm tr}(\rho^2) \leq 1.$ Likewise for $B$.
      Since $C_\rho(A,A^{\dag})C_\rho(B,B^{\dag})\geq |C_\rho(A,B^{\dag})|^2$,
      it follows that $|C_\rho(A,B)|^2 \leq 1.$
\end{enumerate}
More particular cases arise when the system is prepared in a pure state 
$|\psi\rangle$, so that $\rho$ becomes a projection operator and 
$C_{\rho}(A,B)$ reduces to
\begin{eqnarray}
 C_{\rho}(A,B)&\rightarrow&\frac{1}{2}\langle\psi|\{A,B\}|\psi\rangle -
\langle\psi|A|\psi\rangle \langle\psi|B|\psi\rangle =
\frac{1}{2}\langle\{A-\langle A\rangle,B-\langle B\rangle\}\rangle\nonumber\\
& = & {\rm cov}_\rho(A,B),\quad {\rm when~~}[A,B]=0.
\end{eqnarray}
Thus
\begin{equation}
C_\rho(A,A)\rightarrow\langle A^2\rangle-\langle A\rangle^2={\rm var}_\rho(A).
\end{equation}
This is in keeping with the familiar variance-covariance inequality:
\begin{equation}
 {\rm var_{\rho}}(A){\rm var_{\rho}}(B) \geq |{\rm cov}_{\rho}(A,B)|^2.
\end{equation}
If $A$ and $B$ are commuting {\em unitary} operators and because 
$|C_\rho(A,B)|^2\leq C_\rho(A,A)C_\rho(B,B)\leq ({\rm tr}(\rho^2))^2 \leq 1,$
we see that alternative covariance only attains a value of 1 for pure states.

It is worthwhile comparing the two covariance functions, in relation to
two commuting observables, $A,B$. Since $[A,B]=0$, select an orthonormal basis
$|i\rangle$ wherein the operators are simultaneously diagonalised, so
$$A=\sum_i |i\rangle a_i\langle i|,\quad  B=\sum_i |i\rangle b_i\langle i|.$$
Then 
\begin{eqnarray*}
{\rm cov}_\rho(A,B)&=&{\rm tr}(\rho AB)-
          {\rm tr}(\rho A){\rm tr}(\rho B)\\
&=& \sum_i a_ib_i\langle i|\rho|i\rangle -
    \sum_{i,j} a_ib_j\langle i|\rho|i\rangle\langle j|\rho|j\rangle\\
&=&\sum_{i,j}a_i(b_i-b_j)\langle i|\rho|i\rangle\langle j|\rho|j\rangle,
\end{eqnarray*}
and similarly
\begin{eqnarray*}
C_\rho(A,B)&=&{\rm tr}(\rho^2 AB)-{\rm tr}(\rho A \rho B)\\
&=&\sum_{i,j}a_i(b_i-b_j)\langle i|\rho|j\rangle\langle j|\rho|i\rangle,
\end{eqnarray*}
since $\sum_j\langle j|\rho|j\rangle={\rm tr}(\rho)= 1$. Furthermore note that
$\sum_{i,j}\langle i|\rho|j\rangle\langle j|\rho|i\rangle={\rm tr}(\rho^2)
\leq 1.$ Upon symmetrising the sums, we obtain the neater expressions,
\begin{equation}
{\rm cov}_\rho(A,B)=\sum_{i,j}(a_i-a_j)(b_i-b_j) \langle i|\rho|i\rangle
                \langle j|\rho|j\rangle /2
\end{equation}
\begin{equation}
C_\rho(A,B)= \sum_{i,j}(a_i-b_j)(b_i-b_j)\langle i|\rho|j\rangle
             \langle j|\rho|i\rangle /2
\end{equation}
Whilst the ordinary covariance has a clear meaning---namely, a measure of 
the correlations between the results of local measurements $A$ and $B$ that 
commute---the interpretation of the alternative covariance is less obvious.

We can obtain more insight by choosing $A=B$. Since $\rho$ is a positive 
definite hermitian operator, $\langle i|\rho|i\rangle\langle j|\rho|j\rangle
\geq \langle i|\rho|j\rangle\langle j|\rho|i\rangle$, for any two states 
$|i\rangle, |j\rangle$. Therefore for a general (mixed state) density matrix,
$$\sum_{i,j}(a_i-a_j)^2 \langle i|\rho|i\rangle\langle j|\rho|j\rangle \geq
  \sum_{i,j}(a_i-a_j)^2 \langle i|\rho|j\rangle\langle j|\rho|i\rangle,$$
or 
\begin{equation}
 {\rm var}_\rho(A) \geq C_\rho(A,A).
\end{equation}

In the light of the variance inequality above it is natural to ask whether
$$ |{\rm cov}_\rho(A,B)| \geq |C_\rho(A,B)| $$
is true. In fact a single (but carefully chosen) counterexample suffices to
show that it is false: in local bases $u,d$ for two local operators 
$A\otimes 1$ and $1\otimes B$, take 
$$\rho = |(uu+dd)\rangle\langle (uu+dd)|/4 + |ud\rangle\langle ud|/4 +
         |du\rangle\langle du|/4 $$
and select the local operators to be diagonal,
\begin{eqnarray*}
A\otimes 1 &=& (|uu\rangle\langle uu| + |ud\rangle\langle ud|) - 
                (|du\rangle\langle du| + |dd\rangle\langle dd|)\\
1\otimes B &=& (|uu\rangle\langle uu| - |ud\rangle\langle ud|) - 
                (|du\rangle\langle du| - |dd\rangle\langle dd|).
\end{eqnarray*}
Evaluation of the two types of covariance leads to  
$${\rm cov}_\rho(A,B)=0, \quad {\rm but} \quad  C_{\rho}(A,B) = 1/4. $$
Thus the variance inequality cannot be extended to covariance.

However, an immediate consequence of the inequality, var$_\rho(A)\!\geq\! 
C_\rho(A,A)$, is that when var$_\rho(A) =0$, $C_\rho(A,A) =0$ too for
any observable $A$. But $2C_\rho(A,A) = {\rm tr}([A,\rho][A,\rho]^{\dag})$;
so $[A,\rho]=0$, which means that $\rho$ is purely in an eigenstate of $A$.
This accords with the basic tenets of quantum mechanics of course. The 
contrapositive of this result is that if $[A,\rho]\neq 0$, then 
${\rm var}_\rho(A)>0$.

Another worthwhile comment stems from the observation that if $X$ is 
conjugate to $A$ in the sense $[A,X] = i\hbar$, then
$$2C_\rho(A,A)={\rm tr}\left([\rho,A][\rho,A]^{\dag}\right)=\hbar^2{\rm tr}
 \left(\frac{\partial \rho}{\partial X}\frac{\partial \rho}{\partial X}^{\dag}
  \right).$$
Thus,
$${\rm var}_\rho(A)= (\Delta A)^2 \geq \hbar^2{\rm tr}\left(\left|
                     \frac{\partial \rho}{\partial X}\right|^2\right)/2,$$
with equality only applying to pure states. For example, the energy 
uncertainty is given by $(\Delta H)^2 \geq \hbar^2{\rm tr}
\left(\left|\frac{d \rho}{dt}\right|^2\right)/2$, while the momentum uncertainty is given
by the derivative of the density matrix with respect to position: 
$(\Delta P)^2 \geq \hbar^2\;{\rm tr}\left(\left|
   \frac{\partial \rho}{\partial X}\right|^2\right)/2$, and so on.

For a general mixed configuration, the two inequalities,
$${\rm var}_\rho(A){\rm var}_\rho(B) \geq C_\rho(A,A)C_\rho(B,B)
   \geq \left| C_\rho(A,B)\right|^2 \]
together with the well-known
$${\rm var}_\rho(A){\rm var}_\rho(B) \geq |{\rm cov}_\rho(A,B)|^2,$$
provide a lower bound for the experimentally observed variance products of 
any two operators, whether or not they commute. For instance,
$${\rm var}_\rho(X).{\rm var}_\rho(P) \geq \left|\half\hbar^2 {\rm tr}\left(
 \frac{\partial\rho}{\partial X}\frac{\partial\rho}{\partial P}\right)
 \right|^2.$$
In the next section we present
examples of operators $A,B$, for which there exist states such that 
$|{\rm cov}_\rho(A,B)| \geq |C_\rho(A,B)|$ and also other states for which 
$|C_\rho(A,B)| \geq |{\rm cov}_\rho(A,B)|$. Hence {\em both} inequalities 
must be considered jointly in an examination of the minimum of the variance 
product, together with Heisenberg's well-known lower bound,
$|{\rm tr}(\rho [A,B])|/2$.

\section{Correlation measures for pure states of two subsystems}

This section examines the correlation properties of the entanglement of two 
subsystems in a tensor product Hilbert space ${\Bbb H}^{(1)}\otimes 
{\Bbb H}^{(2)}$. By definition, measurements can be carried out without 
mutual interference on the two subsystems so their corresponding observables, 
superscripted by (1) and (2), are commuting operators. As mentioned in the
introduction, a state of the system is factorisable or disentangled if the 
larger density matrix $\rho$ is merely a direct product of density matrices 
associated with the two subsystems, or $\rho = \rho^{(1)} \otimes \rho^{(2)}$.
For any two local operators $A^{(1)}=A\otimes 1,B^{(2)}=1 \otimes B$, it is 
easy to show that the covariance disappears:
\begin{equation}
 {\rm cov}_{\rho^{(1)}\otimes\rho^{(2)}}(A^{(1)},B^{(2)}) = 
 \langle (A-\langle A\rangle)\otimes\langle(B-\langle B\rangle)\rangle = 0.
\end{equation}
This includes the case of a pure disentangled state, $|\phi\rangle= |\phi^{(1)}
\rangle |\phi^{(2)}\rangle$.

Having noted that the covariance is non-zero in disentangled states, we now 
refer to the conditions imposed upon any measure of entanglement. The second 
condition is that it be invariant under local unitary transformations. With 
this in mind, define the {\em covariance entanglement} for pure states as 
$$ E_{A^{(1)},B^{(2)}}(\rho) \equiv \max_{U=U^{(1)}\otimes U^{(2)}} 
|{\rm cov}_{U\rho U^{\dagger}}(A^{(1)},B^{(2)})|.$$
The maximum will clearly be invariant under additional local unitary 
transformations.

Since the operation of permuting the elements of a Hilbert space is unitary,
all elements of the Hilbert spaces are equally important. For this reason, 
it is natural to select the operators $A^{(1)}$ and $B^{(2)}$ so as to 
equally weight the elements of the Hilbert space. The next section describes 
several ideas for achieving this, starting with the simplest case.

\subsection{Pure state correlations for $N^{(1)}=N^{(2)}=2$}
This section investigates  a method for quantifying pure state entanglement
in the simplest possible non-trivial case, corresponding to two spin 1/2
systems, with Hilbert space ${\Bbb H} \otimes {\Bbb H}$, where ${\Bbb H} =
{\Bbb C}^2$ is a local Hilbert space, with orthonormal basis $|u\rangle,
|d\rangle$. Consider two local operators which distinguish between elements of
the local Hilbert spaces. With the aim of weighting local basis elements 
equally, define operators in the product basis $|uu\rangle,|ud\rangle,
|du\rangle,|dd\rangle$ by
$$A^{(1)}=\sigma_3^{(1)}=\!\left(\begin{array}{cccc}
           1 &  &  & \\
             & 1 &  & \\
             &  & -1 & \\
             &  &  & -1
           \end{array} \right),
 B^{(2)}=\sigma_3^{(2)}=\!\left(\begin{array}{cccc}
           1 &  &  & \\
             & -1 &  & \\
             &  & 1 & \\
             &  &  & -1
           \end{array} \right),$$
and so
$$A^{(1)} B^{(2)}=\sigma_3^{(1)}\sigma_3^{(2)} =
       \!\left(\begin{array}{cccc}
           1 &  &  & \\
             & -1 &  & \\
             &  & -1 & \\
             &  &  & 1
           \end{array} \right).$$

\begin{figure}[tb]
{\centering
\mbox{\centering \epsfxsize=8cm \epsfbox{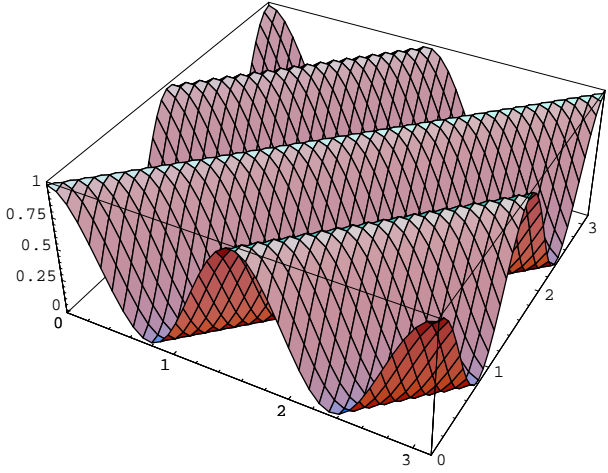}}}
\caption{\label{covariancefigure} Equal-weight covariance of the Bell-state 
$|uu+dd\rangle/\sqrt{2}$ under equal local unitary transformations about two 
axes.}
\end{figure}

Next consider the pure (normalized but arbitrary) state,
$$|\phi\rangle = \alpha|uu\rangle + \beta|ud\rangle +
                 \gamma|du\rangle + \delta|dd\rangle.$$
Working out cov$_\rho(A,B)$ in this state, it is straightforward to show that 
the covariance is maximised provided that $|\alpha|=|\delta|=1/\sqrt{2},
\beta=\gamma=0$, or $|\beta| = |\gamma| = 1/\sqrt{2}, \alpha = \delta = 0$. 
Thus one may take the four independent Bell states,
$$|1 0\rangle \equiv [|ud\rangle + |du\rangle]/\sqrt{2},\quad
  |0 0\rangle \equiv [|ud\rangle - |du\rangle]/\sqrt{2},$$
$$|1 +\rangle \equiv [|uu\rangle + |dd\rangle]/\sqrt{2},\quad
  |1 -\rangle \equiv [|uu\rangle - |dd\rangle]/\sqrt{2},$$
as the ones that have the largest covariance. (These pure states are also 
known to be the most entangled ones.) Of course they are all local unitary 
transforms of just one of them, say the Bell state, 
$|1 +\rangle \equiv |(uu + dd)\rangle/\sqrt{2}$,
with a corresponding $\rho =|uu+dd\rangle\langle uu+dd|/2$. If one rotates
the operators $A,B$ {\em together} about the ``$y$-axis'' by the same amount,
we can get a good idea of how the covariance varies with rotation angle; 
maximization is attained when the angle is $n\pi$. See Figure
\ref{covariancefigure}.

\begin{figure}[tb]
{\centering \mbox{\centering \epsfxsize=8cm \epsfbox{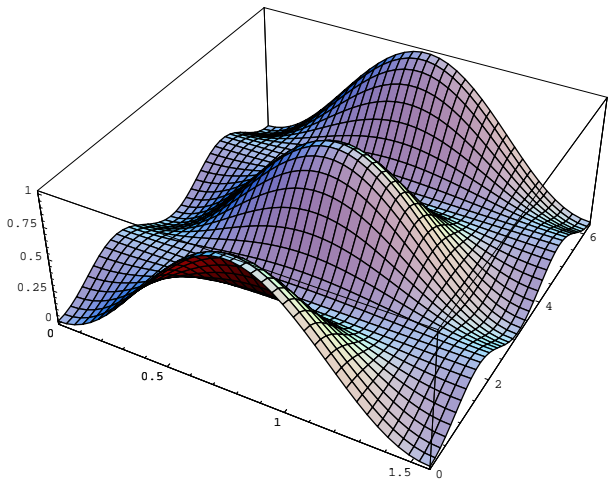}}}
\caption{\label{variancefigure}Equal-weight variance of a parametrised set of 
pure states, under different relative orientations of the state and the local 
operators. The maximally entangled states have $x=\pi/4$, and are the only 
states to attain a covariance of 1, whilst the disentangled states 
($x=0,\pi/2$) are the only ones which attain a covariance of zero, for 
particular relative transformations of the local operator.}
\end{figure}

If a pure state is disentangled, then there is an orientation of the
local $A^{(1)}$ which has zero variance. To see this, rotate the state by 
local unitary transformations until the reduced density matrix for the local 
operator in question is diagonalised. Since the initial state was pure and 
disentangled, then it may be represented by a separable projector 
$\rho=\rho^{(1)}\rho^{(2)}$, in which both reduced density matrices are 
projectors. Thus the main diagonals of $\rho^{(1)}$ and $\rho^{(2)}$ can be 
reduced to a single 1, with 0s elsewhere. Hence for diagonalised local 
operators $A$ and $B$ in this  basis, 
$$\langle A^2\rangle = \langle A\rangle^2,\quad\langle B^2\rangle = \langle 
B\rangle^2,$$
so the variances vanishes. Figure \ref{variancefigure} illustrates the 
behaviour of the variance in a 2-variable parametrisation. 
The horizontal axis variable $x$ parametrises a set of pure states which 
range from disentangled to a maximally entangled Bell
state, and back to disentangled again, i.e. $\rho=|\psi(x)\rangle\langle
\psi(x)|$ where $|\psi(x)\rangle = \cos(x)|uu\rangle+\sin(x)|dd\rangle$.
The second variable $y$ parametrises $y$-axis rotations of the local spin
basis (1) associated with $A$ alone.

These results may be applied to actual spin measurements. If one knows that 
a state is pure, but is not certain of the degree of entanglement, local spin 
measurements can be made in a variety of directions. If the variance and 
covariance of these measurements vanishes in a pure state, then the state 
must be disentangled.

\subsection{Pure state correlations for $N^{(1)}=N^{(2)}\equiv N >2$}
Many different choices of local operators $A^{(1)},B^{(2)}$ are possible, and 
different choices will lead to different behaviour of the covariance. However
before considering two sensible choices, let us note that for $N^{(1)}=N^{(2)}
=2$, one of the maximally entangled states can be taken to be the state of
total spin $|10\rangle$, while minimally entangled states are $|11\rangle,
|1-1\rangle$. Now for these state combinations maximal entanglement happens
to equate with maximal dispersion $(\Delta J)^2$ and zero entanglement equates
with minimal dispersion $(\Delta J)^2$, where {\bf $J$} stands for total
spin. This suggests that for higher spin, some maximally entangled states 
might be found by minimising the total angular momentum dispersion and vice 
versa. This approach towards quantifying entanglement is quite interesting in 
its own right and is pursued in Appendix A, where we also tie it to the 
Majorana-Penrose pictorial view of spin. In Appendix B, by contrast, we
classify any measure of entanglement via an integrity basis for density
matrix invariants.

\subsubsection{Pair-discrimination}
Consider two-system Hilbert spaces where the local spaces may each have
dimension greater than 2. Select local operators $A^{(1)} = A\otimes1, 
B^{(2)}=1\otimes B$, where $A,B$ may be expressed in their respective local
bases as unitary transformations of the following diagonal matrix,
\begin{equation}
\left( \begin{array}{cccc}
        1 & 0 & 0 & \cdots\\
        0 & -1 & 0 & \cdots\\
        0 & 0 & 0 & \cdots\\
        \cdots & \cdots & \cdots &\ddots
       \end{array} \right).
\end{equation}
Next, maximise the covariance over all such unitarily-transformed matrices; 
the result is perforce invariant under local unitary transformations of the 
state. Labelling the local bases $|a_i\rangle$ and $|b_i\rangle$ respectively,
where $i$ runs from 1 to $N$, operators like the above discriminate between 
pairs of elements in a local subspace of the full Hilbert space, and 
treat terms of the form $|a_ib_k+a_jb_l\rangle$ as the basic element of 
entanglement\footnote{Another possibility is to replace all the zeroes 
along the main diagonal of $A$ or $B$ with $1$ or $-1$; this makes the 
operator $U$ unitary, which means that its variance is simply 
$1 - |\langle U \rangle|^2$. However the distribution of the $\pm 1$ 
eigenvalues is not self-evident, except for the spin $\half \otimes$ 
spin$\half$ case.}.

\subsubsection{Equal weight unitary operators}
Since we wish to handle all the subsystem states democratically, let us
define an equal weight local unitary operator as consisting of some unitary 
transformation of a diagonal matrix comprising the $N$-th roots of unity. 
(Note that these matrices are not hermitian when $N\geq 3$ and cannot
correspond to observables.) Here the local weight unitary matrices in their 
corresponding diagonalising basis, up to an overall phase, are given by
$$\left( \begin{array}{cccc}
         \exp(2i\pi/N) & & & \\
            & \exp(4i\pi/N) & & \\
            &  & \ddots & \\
            &  &   & 1 \end{array}\right).$$
Only for the spin $\half$ $\times$ spin $\half$ case, do these operators
correspond exactly to the pairwise local unitary operators used previously.
The next nontrivial case is $N=3$, or spin 1 $\times$ spin 1. In this case
one can see that all states which are local unitary transformations of the 
pure state $|a_1b_1+a_2b_2+a_3b_3\rangle/\sqrt{3}$ have a maximal covariance 
of 1. To understand why, rotate the equal-weight unitary operators so that 
the local states $a_1,a_2,a_3$ and $b_1,b_2,b_3$ produce eigenvalues which 
are respectively conjugate pairs. This yields $\langle A^{(1)}B^{(2)}\rangle
= 1$. However, $\langle A^{(1)} \rangle = \langle B^{(2)}\rangle = 0$, so the
maximised covariance is 1.

What of other states, such as $|a_1b_1 + a_2b_2\rangle$ when $N=3$? The 
following theorem gives a necessary and sufficient condition for a state 
to exhibit $C_\rho=1$, with respect to these equal weight operators and is in 
agreement with all other pure state entanglement measures.
 
\noindent
{\bf Theorem:} {\em The only pure states which attain the maximised covariance 
of 1 under equal-weight local operators are states which are local unitary 
transformations of $|a_1b_1+a_2b_2+\ldots+a_Nb_N\rangle /\sqrt{N}$. 
Any other states exhibit smaller correlations.} 

\noindent
{\bf Proof:} In a tensor product space ${\Bbb H}^{(1)}\otimes{\Bbb H}^{(2)}$, 
consider the state
$$ |\psi\rangle = \sum_{i,j} c_{ij} |a_ib_j\rangle, $$
where $p_{ij}\equiv |c_{ij}|^2 \geq 0$ and $\sum_{i,j}p_{ij}=1$, and the 
orthonormal bases $|a_i\rangle,|b_j\rangle$ are a complete set of eigenstates 
for the diagonalised equal-weight unitary operators $A^{(1)},B^{(2)}$, with 
eigenvalues $a_k=\exp(2i\pi k/N),b_j=\exp(2i\pi j/N)$ respectively. 
Recalling the result for such $A^{(1)},B^{(2)}$ that
$$(1-|\langle A^{(1)}\rangle|^2)(1-|\langle B^{(2)}\rangle|^2) \geq |\langle 
A^{(1)} B^{(2)}\rangle-\langle A^{(1)}\rangle\langle B^{(2)}\rangle|^2, $$
we see that a maximised covariance of 1 is only attainable in a state where 
$\langle A\rangle=\langle B\rangle=0$, whereupon the covariance reduces to
$$C_\psi(A^{(1)},B^{(2)})=\left|\sum_{i,j}p_{ij}a_ib_j\right|.$$
In what cases is this expression maximised, subject to the condition that the 
mean values of the operators remain zero? 

We are seeking to maximise $|\sum p_{ij} a_ib_j|$ subject to 
$$\sum p_{ij}=1,\quad |\sum p_{ij}a_i|=0, \quad|\sum p_{ij} b_j|=0.$$
By the triangle inequality,
$$|\sum p_{ij} a_ib_j| \leq p_{11}|a_1b_1| + |\sum_{i,j>1}p_{ij}a_ib_j|
\leq \ldots \leq \sum_{i,j}p_{ij}|a_ib_j|, $$
where equality holds at every stage only if all the complex numbers $a_i,b_i$
have equal and opposite phase. This means that we are pairing $a_i,b_j$ such 
that non-zero $p_{ij}$ (only for $i=j$) are associated in a one-to-one manner 
with $a_ib_i=1$ for all such pairs, otherwise the parallelism of the complex 
numbers will be lost. This will ensure that
$$\sum_{i}p_{ii}|a_ib_i| = \sum_{i}p_{ii} = 1.$$
At the same time we have to guarantee that the average values of $A$ and $B$
vanish or $|\sum p_{ii} a_i|=0, |\sum p_{ii} b_j|=0$. Since $\sum_i a_i =
\sum_i{b_i} = 0$, a {\em sufficient} condition for this is that for all such 
pairs, the weightings are equal or $p_{ii}=1/\sqrt{N}$; in other words every 
$a_i$ and its corresponding $b_i=a_i^*$ only occurs at most once in the terms 
with equal non-zero weighting. (Actually one may introduce an arbitrary phase
into $c_{ii}$ without affecting this conclusion, but we have chosen not to do 
so.)

Having established that the states which maximise the covariance can take the 
form $|a_ia_1^*+a_2a_2^*+\ldots\rangle$, we should point out that it is not 
necessary for {\em all} the terms to be paired up. Consider the case $N=4$, 
or spin $3/2 \otimes$ spin $3/2$ systems, with bases $a_1,\ldots,a_4$ and 
$b_1,\ldots,b_4$ respectively. It is possible to attain maximised covariances 
of 1 under the equal-weight measures {\em both} for
$|\psi\rangle=|a_1b_1+a_2b_2\rangle/\sqrt{2}$ and for 
$|\phi\rangle=|a_1b_1+a_2b_2+a_3b_3+a_4b_4\rangle/2$. This is achieved by 
choosing eigenvalues so that $\langle A^{(1)}\rangle=\langle B^{(2)}\rangle=0$ 
and yet pick eigenvalues such that the values of $A^{(1)}B^{(2)}$ in the 
states $|a_1b_1\rangle\ldots |a_4b_4\rangle$ are all 1. As we are in a 
4-dimensional space, we can achieve this by taking the eigenvalue sets 
$\{1,-1\}$ and $\{1,i,-1,-i\}$ for both operators on both states 
$|\psi\rangle$ and $|\phi\rangle$ respectively.

This observation means that the `equal-weight' $C$-based measures for $N>2$
are not measures of entanglement, under the standard criteria. 
Information-based entanglement measures, such as the relative entropy, specify
that $|\psi\rangle$ is less entangled than $|\phi\rangle$. It appears that 
the covariance-based measures of entanglement are most useful when dealing 
with spin $\half \otimes$ spin$\half$ systems, since in this case there is no 
ambiguity as to the choice of eigenvalues. Similarly for operators of
prime dimension, such ambiguities are absent, because there is only one way
to arrive at a maximally correlated state: the non-uniqueness only pertains
to composite-dimensional local spaces.

The above result demonstrates that for pure states, the reduced density 
matrices must be diagonalised in order to maximise the covariance of the 
diagonalised local unitary operators. However, for mixed states it is not at
all obvious that the reduced density matrices must be diagonalised in order 
to maximise the unitary matrix covariance. This issue will will be examined 
in section 4.

\subsection{Pure state correlations for $N^{(1)}\neq N^{(2)}$}

For spaces of differing dimension, such as spin $1\times$ spin $1/2$, the 
covariance of the pairwise operators behave just as in the other cases. 
However, if the equal-weight local operators are used, covariances of 1 are 
{\em not attainable}. This reflects the fact that the bases have different 
sizes, and so there is no way to pair up elements between the bases in a 
one-to-one manner so as to produce a set of product eigenvalues with the same 
phase.

The simplest example which exhibits this effect is a spin $\half\;\otimes$ 
spin $1$ space, with equal-weight operators
\begin{equation}
A^{(1)}={\rm diag}(1,-1)\otimes 1,\quad B^{(2)}=1\otimes{\rm diag}
(1, \exp(2i\pi/3), \exp(4 i\pi/3))
\end{equation}
As with the proof that the states of maximum covariance are 
$|a_1b_1+\ldots+a_nb_n\rangle$, a covariance of 1 is only attainable if 
$\langle A^{(1)}B^{(2)}\rangle=1$; we also need the complex numbers 
$p_{ij} a_ib_j$ to have the same phase (where the state has Schmidt 
decomposition $|\phi\rangle=\sum_{ij}\sqrt{p_{ij}}|a_ib_j\rangle$ in the 
basis of eigenvalues of the operators $A^{(1)},B^{(2)}$). Following similar 
arguments to those used in the proof, it is not possible to choose $p_{ij}$ 
so that the direction condition is satisfied; this is because no repetitions 
of $a_i$ or $b_j$ values can occur in the set of non-zero $p_{ij}$, since
the resulting $p_{ij}a_ib_j$ would not have the same phase. But it is not 
possible to partially pair up the given set of eigenvalues so that the 
directions of the products are the same, by straightforward enumeration of 
the cases. Thus states in this basis cannot attain covariances of $1$.

Clearly many other local operators can be defined which provide a variety 
of different correlation measures for two-system states but none stands out.

\section{Correlation measures for mixed states}
Making a distinction between quantum and classical correlations has proved a 
thorny problem in the study of quantum entanglement. The nature of the problem
may be seen when comparing the two states, one pure and one mixed, which 
possess the same covariance for $A^{(1)}=B^{(2)}=\sigma_3$:
$$\half|uu+dd\rangle\langle uu+dd| \quad {\rm and}\quad
   \half|uu\rangle\langle uu| + \half|dd\rangle\langle dd|.$$
The first state is a Bell state and is maximally entangled, whilst the second 
state is a mixture of disentangled projectors, and is normally regarded 
as being disentangled. As both states exhibit correlations, it is natural to 
ask whether the alternative covariance introduced in section 2 provides a way 
of distinguishing between classical and quantum correlations.

\subsection{Distinction between $C_\rho$ and cov$_\rho$}

Take any two commuting local measurements, $A^{(1)}=A\otimes 1, 
B^{(2)}=1\otimes B$, and define the function
\begin{equation}
 E_{AB}(\rho) \equiv {\rm max}_U|C_{U\rho U^{\dagger}}(A^{(1)},B^{(2)})|^2,
\end{equation}
where the maximum is now taken over all local unitary transformations of the
general {\em mixed} density matrix $\rho$. In previous sections we 
investigated the behaviour of this function for pure states (when $C_\rho$ 
reduces to the covariance), and found that it appeared to have many of the 
properties desirable in a pure state entanglement measure.

The situation where the density matrix corresponds to an impure configuration
is more intriguing. Shown below is a comparison of the behaviour of the two 
maximised covariance entanglement measures, in several example configurations,
which illustrate the distinction between 
$|{\rm cov}_{\rho}(A^{(1)},B^{(2)})|^2$ and 
$|C_{\rho}(A^{(1)},B^{(2)})|^2$.
\vspace{.2in}

\noindent
\begin{tabular}{lcc}
 \hline
 Density matrix & $|{\rm cov}_{\rho}(A^{(1)}\!,\!B^{(2)})|^2$ & 
 $|C_{\rho}(A^{(1)}\!,\!B^{(2)})|^2$\\ \hline
 $\rho_1=[|uu\rangle\langle uu|+|dd\rangle\langle dd|]/2$ & 1 & 0 \\
 $\rho_2=[\half|uu\rangle\langle uu|\!+\!\frac{1}{4}|uu\!+
  \!dd\rangle\langle uu\!+\!dd|]$  & 3/4 & 1/4 \\
 $\rho_3=|uu\rangle\langle uu|$ & 0 & 0\\
 $\rho_4=|uu\!+\!dd\rangle\langle uu\!+\!dd|/2$ & 1 & 1 \\ \hline
\end{tabular}
\vspace{.2in}

The illustrative state $\rho_4$ is pure but entangled, $\rho_3$ is
factorizable and therefore
disentangled, while the matrix $\rho_1$ is not factorizable but can be
expressed as a sum of separable projectors; therefore $\rho_1$ should 
represent a disentangled configuration, according to standard expectations. 
By inspecting the table we see that cov$_{\rho}(A^{(1)},B^{(2)})$, being 
nonzero, is not a good entanglement measure $E(\rho)$, but the alternative 
$C_{\rho}(A^{(1)},B^{(2)})$ is better in that it does vanish.

The alternative covariance $C$ in mixed states is bounded above by  
${\rm tr}(\rho^2)$, so the state must 
be pure to obtain a covariance of 1 under local unitary transformations. 
Another point worth remembering is that for configurations of maximum entropy 
where $\rho \propto 1$ and commutes with all operators, $C$ is automatically
zero. 

\begin{figure}[p]
(i) $\left|{\rm cov}_\rho(A^{(1)},B^{(2)})\right|$

{\centering \mbox{\centering \epsfxsize=8cm 
\epsfbox{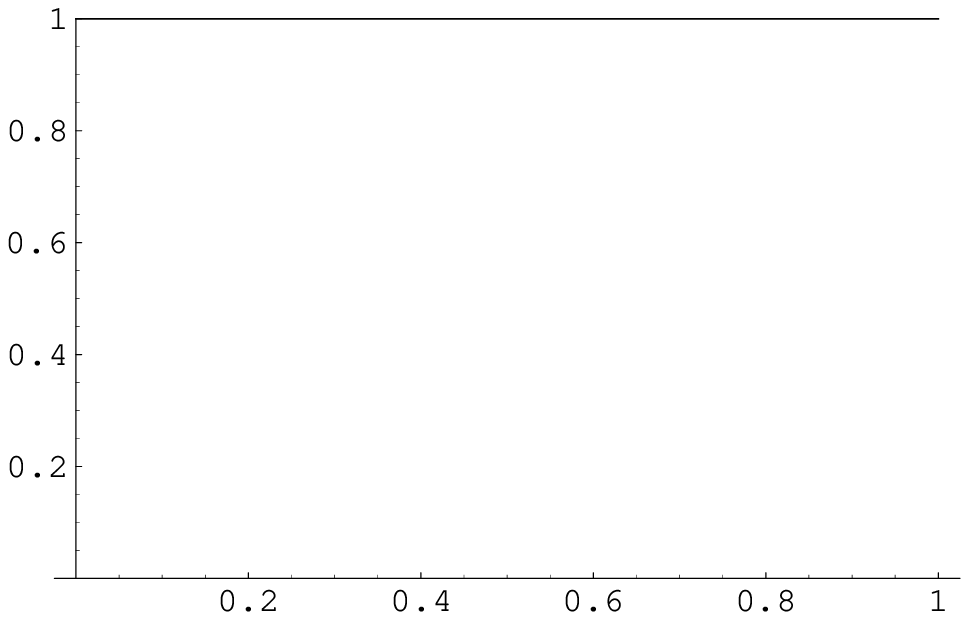}}}

(ii) $\left|C_{\rho}(A^{(1)},B^{(2)})\right|$

{\centering \mbox{\centering \epsfxsize=8cm 
\epsfbox{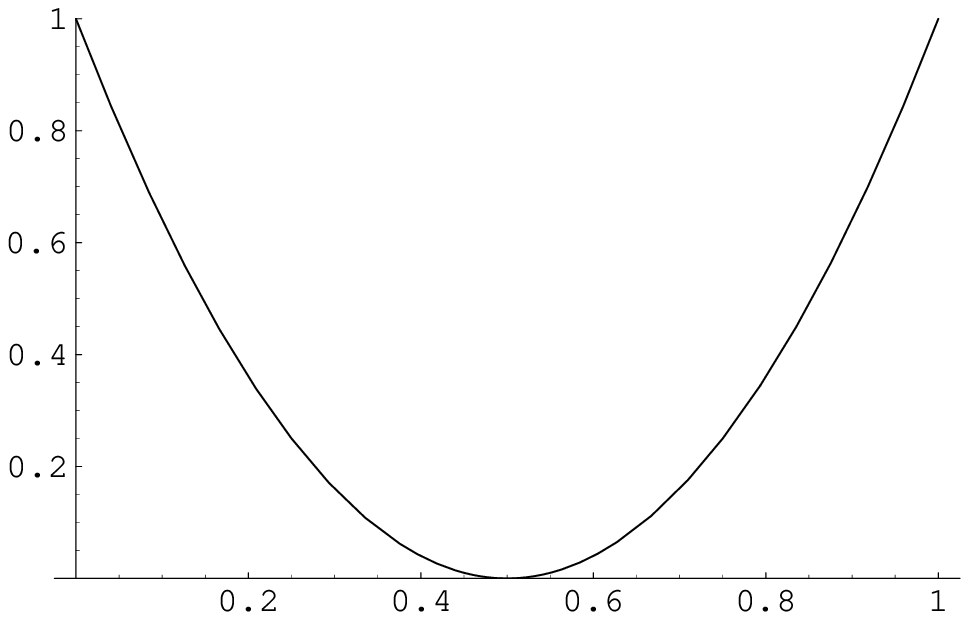}}}

\caption{\label{mixedparametrisation1}(i) Equal-weight covariance magnitude 
of a parametrised mixture of different Bell-states. (ii) Equal weight 
alternative covariance magnitude for the Bell-state mixture.
The parametrisation in both cases is any one of the following: 
$\rho = x|b_1\rangle\langle b_1|+(1-x)|b_2\rangle\langle b_2|,$
where either $|b_1\rangle=|ud\pm du\rangle/\sqrt{2}$, $|b_2\rangle=|ud\mp du
\rangle/\sqrt{2}$, or else $|b_1\rangle=|uu\pm dd\rangle/\sqrt{2}$,
$|b_2\rangle=|ud \mp du\rangle/\sqrt{2}$. The Bell-states must be different 
in these parametrisations, to obtain a non-trivial result. The result is not
a maximisation over all local unitary transformations.}
\end{figure}

\begin{figure}[p]
(i) $\left|{\rm cov}_\rho(A^{(1)},B^{(2)})\right|$

{\centering \mbox{\centering \epsfxsize=8cm 
\epsfbox{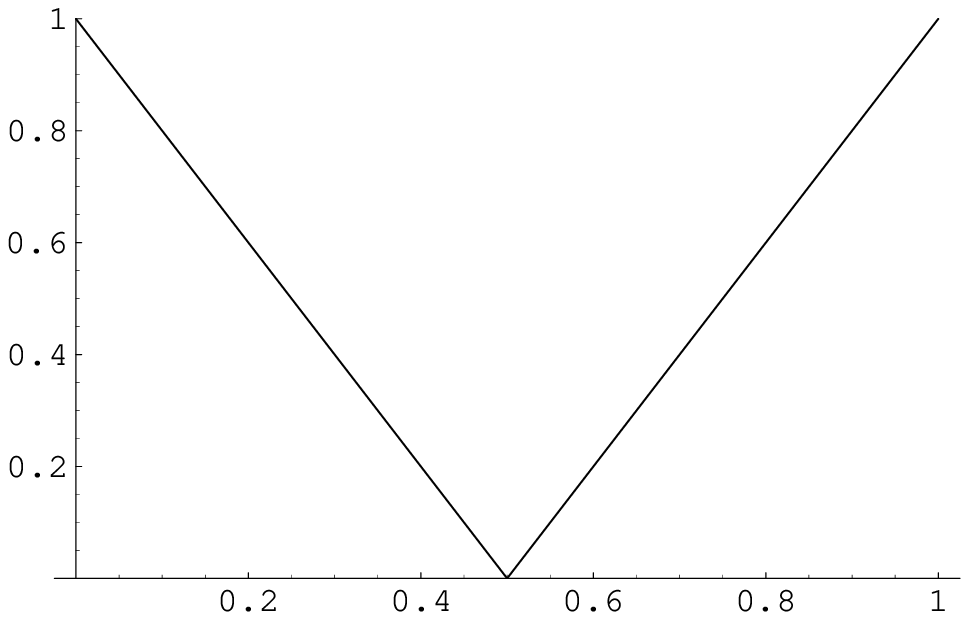}}}

(ii) $\left|C_{\rho}(A^{(1)},B^{(2)})\right|$

{\centering \mbox{\centering \epsfxsize=8cm 
\epsfbox{bell.eps}}}

\caption{\label{mixedparametrisation2}(i) Equal-weight covariance magnitude 
of a parametrised mixture of different Bell-states. (ii) Equal weight 
alternative covariance magnitude for the Bell-state mixture. The 
parametrisation in both cases is \mbox{$\rho=x|b_1\rangle\langle b_1|+(1-x)
|b_2\rangle\langle b_2|,$} where $|b_1\rangle=|uu\pm dd\rangle/\sqrt{2},
|b_2\rangle=|ud\pm du\rangle/\sqrt{2}$. The Bell states in these 
parametrisations may be used in any combination allowed by the above 
expression (i.e. all four sign combinations are permitted). The result is not 
a maximisation over all local unitary transformations.}
\end{figure}

Figures \ref{mixedparametrisation1} and \ref{mixedparametrisation2} depict 
the squares of the equal weight classical covariance and alternative 
covariance of a parametrised mixture of different Bell states. They show that 
both covariance measures attain a maximum of 1 only for the pure states, and 
that the behaviour of these functions depends upon the kind of Bell-states 
involved in the mixtures.

For {\em any} choice of $A^{(1)},B^{(2)}$, a maximised 
$C_{\rho}(A^{(1)},B^{(2)})$ of zero occurs for many mixed configuration $\rho$
which are disentangled, according to the standard definition of a mixed state,
such as the states of maximum entropy with $\rho \propto 1$. This behaviour of
the alternative covariance for `partially entangled' configurations is what 
one would naively expect for a mixed configuration entanglement measure 
$E(\rho)$. However, there is a particular class of mixed states which are 
disentangled according to the commonly accepted definition of a mixed state, 
whilst exhibiting non-zero alternative covariance properties. 

\subsection{Conditions on entanglement measures}
This section assesses the alternative covariance as a measure of entanglement, 
according to the principles outlined by Vedral and Plenio\cite{entanglement}.

Firstly, consider the cases where a state exhibits zero quantum correlation. 
A mixed configuration is usually defined to be separable if it can be written 
as a sum of separable projectors; thus
$$\rho = \sum_j p_j\rho^{(1)}_j\otimes \rho^{(2)}_j;\qquad 
 \rho_i^{(n)}\rho_j^{(n)} = \delta_{ij}\rho_i^{(n)}.$$
Then for any such $\rho$,
\begin{eqnarray*}
C_{\rho}(A^{(1)},B^{(2)}) &=& {\rm tr}(\rho^2 A^{(1)} B^{(2)} - 
          \rho A^{(1)} \rho B^{(2)})\\
&=& {\rm tr}\left(\left(\sum p_i\rho^{(1)}_i A\otimes \rho^{(2)}_i\right) 
     \left(\sum p_j\rho^{(1)}_j \otimes B\rho^{(2)}_j\right)\right) \\
& & -{\rm tr}\left(\left(\sum p_i\rho^{(1)}_i A\otimes \rho^{(2)}_i\right)
 \left(\sum p_j\rho^{(1)}_j \otimes\rho^{(2)}_j B\right) \right) \\
&=&\sum\sum p_ip_j{\rm tr}(\rho^{(1)}_iA\rho^{(1)}_j\otimes\rho^{(2)}_i
    B\rho^{(2)}_j \\
&&\qquad\qquad\qquad -\rho^{(1)}_iA\rho^{(1)}_j\otimes
           \rho^{(2)}_i\rho^{(2)}_jB)\\
&=& \sum \sum p_i p_j{\rm tr}_{(1)}(\rho^{(1)}_iA\rho^{(1)}_j){\rm tr}_{(2)}
     (\rho^{(2)}_iB\rho^{(2)}_j-\rho^{(2)}_i\rho^{(2)}_jB)\\
&{\rm or}&\sum\sum p_i p_j {\rm tr}_{(1)}(\rho^{(1)}_iA\rho^{(1)}_j-
     \rho^{(1)}_i\rho^{(1)}_j A){\rm tr}_{(2)}(\rho^{(2)}_iB\rho^{(2)}_j), 
\end{eqnarray*}
which is guaranteed to vanish if the $\rho_i^{(n)}$ are projectors.

Now consider cases where the density matrix is a mixture like $$\half
|uu\rangle\langle uu|+\frac{1}{8}|(u+d)(u+d)\rangle\langle(u+d)(u+d)|.$$ 
Here, it can be shown that these mixtures produce non-zero $C_{\rho}(A,B)$, 
because their expansion into projector traces may be used to extract 
off-diagonal elements of observables $A,B$ in suitable local bases. Take
for instance the two local operators on separate spin $\half$ bases,
$A\!=\!B\!=\sigma_2$, which is a unitary transformations of the equal-weight 
matrix, $\sigma_3$ used previously. The terms in the expansion of $C$ above 
are non-vanishing only if non-parallel, non-orthogonal projectors are used. 
Thus we evaluate the two non-vanishing terms, with local projectors 
$|u+d\rangle\langle u+d|/2,\;\;|u\rangle\langle u|$ giving
$$\langle u+d|u\rangle\langle u|A|u+d\rangle (\langle u|B|u+d\rangle
 \langle u+d|u\rangle - \langle u+d|B|u\rangle\langle u|u+d\rangle)$$
$$\qquad\qquad\qquad = (-i)(-i - i) = -2,$$
$$\langle u|u+d\rangle\langle u+d|A|u\rangle (\langle u+d|B|u\rangle
 \langle u|u+d\rangle - \langle u|B|u+d\rangle\langle u+d|u\rangle)$$
$$\qquad\qquad\qquad = (i)(i + i) = -2.$$
These terms do not cancel, and the other two terms in the expansion are zero, 
so the alternative covariance $C_\rho$ is non-zero even though we are dealing
with a disentangled state, according to the usual terminology.

From this we deduce that the measure $E$ based on alternative covariance is 
{\em not} a measure of entanglement (according to the conditions provided by 
Vedral and Plenio), when mixed states are encountered, since it violates the 
first condition for such measures. 

\subsection{Local purification procedures}
The third condition on entanglement measures proposed by Vedral and Plenio 
is that the entanglement of a state should not increase under the three types 
of purification processes (LGM, CC, and PS).

Let us consider all possible LGM, CC and PS  measurement operations which act 
on a disentangled state
$$\rho_0=|uu\rangle\langle uu|$$ 
and yield states which exhibit non-zero quantum covariance 
$C_{\rho}(A^{(1)},B^{(2)})$, such as 
$$\rho_1=\frac{1}{2}|uu\rangle\langle uu|+\frac{1}{8}
      |(u+d)(u+d)\rangle\langle(u+d)(u+d)|$$
The change from $\rho_0$ to $\rho_1$ may be performed by the classically 
correlated set of local measurements $V_1\ldots V_8$, where
\begin{eqnarray*}
V_1&=&  \frac{1}{\sqrt{2}}|uu\rangle\langle uu|=\frac{1}{\sqrt{2}}
 |u\rangle\langle u|\otimes |u\rangle\langle u| \\
V_2&=& \frac{1}{2\sqrt{2}} |(u+d)(u+d)\rangle\langle uu| = 
    \frac{1}{2\sqrt{2}}|(u+d)\rangle\langle u|\otimes|(u+d)\rangle\langle u|
\end{eqnarray*}
together with $V_3,\ldots,V_8$ which have $\langle ud|,\langle du|,
\langle dd|$ in place of $\langle uu|$. Thus $\sum V_i\rho_0 V_i^{\dagger}=
\rho_1$, and $\sum V_i^{\dag} V_i \propto 1$, so this represents a complete 
measurement.

Because this set of operations $V_i$ is local, we would expect any measure of 
entanglement to not increase under these operations. However, the state which 
results from this procedure does in fact exhibit non-zero $C_\rho$, as was 
demonstrated earlier. Therefore we have found a complete local general 
measurement for which the entanglement increases based upon alternative 
covariance. Thus the unitarily maximised $C_\rho$ does not satisfy the 
conditions for a measure of mixed state entanglement proposed by Vedral and Plenio.

\section{Conclusion}
We have shown that for pure two 2-state systems, the maximised
covariance agrees with other measures of entanglement in specifying which
states are disentangled and which are maximally entangled. For subspaces
of larger dimension, the situation is less clear-cut since there are
ambiguities in the process of selecting eigenvalues for the local operators, 
but it is possible to obtain information about the degree of higher-order 
correlations of two subsystems in this way.

The variance and covariance could be used to indicate the best measurements 
to make to detect entanglement of bipartite systems, by locating the unitary 
transformation which produces maximum covariance. However, the problem of 
mixed state separability measures is not resolved by the correlation functions
$C$ and the covariance, as we have used them in this paper. Nevertheless the 
link between alternative variance and the conjugate variable derivatives for 
position, momentum, energy and time is intriguing, and might be extended to 
other quantised models involving conjugate variables. The inequality
${\rm var}(X) \geq C(X,X)$, with equality for pure states, should be
applicable to situations where observations are made upon impure states.

It is interesting to observe that for pure density matrices $\rho$, both the 
average value and the variance of an operator $F$ are symmetrical under the 
interchange of $\rho$ and $F$, when expressed in the form
$$\langle F\rangle = {\rm tr}(\rho F), \; {\rm var}(F) = C_{\rho}(F,F) = 
{\rm tr}(\rho^2 F^2 - \rho F \rho F).$$
They motivate the suggestion that a quantum state $|\psi\rangle$ is best 
represented not as a ket, but as a projection operator $P_{\psi}=|\psi\rangle
\langle\psi|$. Any measurement process on the state is relational in the sense
of Rovelli \cite{relentropy}: the notion of state vector reduction is replaced
by the notion that every new measurement process requires a new Hilbert space 
to be defined, with new operators corresponding to the new state and any 
successive measurement made upon the state. The basic idea is that a 
measurement treats the state information in a comparative sense, with meaning 
only in relation to the operator corresponding to the observable quantity 
being measured. 

The process of state reduction to the eigenstate of some observable under a 
measurement is an {\em apparent one} resulting not from the
action of the {\em operator} corresponding to the measurement, but from 
{\em some other} physical process which occurs in the measurement, such as a 
filtering process of some kind (like the case of a Stern-Gerlach 
experiment), causing {\em later} measurements on the physical system to 
be {\em represented} in a completely new basis by new state and measurement 
operators. The adoption of the projector as a unit of relational information
does not change any predictions of quantum mechanics in terms of average
values\footnote{The paradox of Schroedinger's Cat is resolved by
representing the cat's state and the measurement on an equal footing as   
projection operators in the (relational) basis {\em for that measurement}.   
Traces are taken to predict average real-number values, but at no stage
does one say `the cat is in a state', since the state projection operator 
$P_{\psi}$ only has relational significance in this interpretation.}.

\appendix

\section{The Majorana-Penrose Representation of Symmetrised States}
It is well known that a spin state $|j,m\rangle$ can be obtained by forming a
fully symmetric direct product of $n=2j$ spin 1/2 states. Denoting the spin
1/2 states by $|u\rangle = |1/2,1/2\rangle,\,\,|d\rangle = |1/2,-1/2\rangle$,
the spin $j$ states are given by:
\begin{eqnarray}
 |j,j\rangle &=& |\overbrace{u u\ldots u}^{2j}\rangle\nonumber\\
 |j,j-1\rangle &=& [|du\ldots u+ud\ldots u+\cdots+uu\ldots d\rangle]/\sqrt{2j}
 \nonumber\\
 \cdots & & \cdots \nonumber\\
|j,m\rangle &=& \sum_{\rm sym}|(u)^{j+m}(d)^{j-m}\rangle/\sqrt{^{2j}C_{j+m}}\\
 \cdots & & \cdots\nonumber\\
 |j,-j\rangle &=& |dd\ldots d\rangle\nonumber.
\end{eqnarray}
In the Majorana-Penrose representation \cite{MP} these states are mapped onto 
the unit sphere, with stereographic projection $z$ taken from the South Pole, 
onto the complex plane by making the association:
\begin{equation}
 |j,m\rangle \leftrightarrow \sqrt{^{2j}C_{j+m}}z^{j+m}.
\end{equation}

This association may be rewritten in terms of the spin-$\half$ basis in an 
elegant way which emphasises that the $z^i$ powers are proportional to a 
symmetrised sum where $i$ spin-$\half$ terms of spin $|u\rangle$, and $(n-i)$ 
spin $|d\rangle$ terms are selected, namely
\begin{equation}
^{2j}C_{j+m} z^{j+m}\leftrightarrow 
|u\ldots ud\ldots d\rangle+\ldots+|d\ldots du\ldots u\rangle
\end{equation}
where the sum is taken over all ways of selecting $i$ spin-$\half$ elements 
from the $2j+1$ elements in the spin-$j$ space.

Thus from a general spin state $|\psi\rangle=\sum_m \psi_{jm}|j m\rangle$ we
can define a polynomial
$$p(z) = a_{2j}z^{2j} + a_{2j-1}z^{2j-1} + \cdots + a_01,$$
where $a_{j+m} \equiv \sqrt{^{2j}C_{j+m}}\psi_{jm}z^{j+m}$. The roots of this
polynomial\footnote{The action of exponential functions of angular momentum
operators on Majorana-Penrose polynomials are amusing. We quote them without 
proof: (i) $\exp(\xi J_z).p(z) = \exp(-\xi j)p(z{\rm e}^{\xi});$
(ii) $\exp(\xi J_-).p(z)=p(z+\xi);$ (iii) $\exp(\xi J_+).p(z)=
z^{2j}p(z/(\xi z+1))$.} give $2j$ points on the stereographic plane and,
correspondingly, a general spin $j$ state, being some superposition over $m$ 
values, will be described by $2j$ distinct points on the Poincar\'{e} sphere, 
obtained by stereographic projection of the roots placed on the horizontal 
$x$-$y$ plane. It is worth remarking that another spin state $|\psi'\rangle$, 
represented by another polynomial,
$$p'(z) = a'_{2j}z^{2j} + a'_{2j-1}z^{2j-1} + \cdots + a'_01,$$
yields a scalar product,
$$\langle\psi'|\psi\rangle = \sum_m a_{j+m}a'^*_{j+m}/^{2j}C_{j+m}.$$

\begin{figure}[tb]
{\centering \mbox{\centering \epsfxsize=8cm \epsfbox{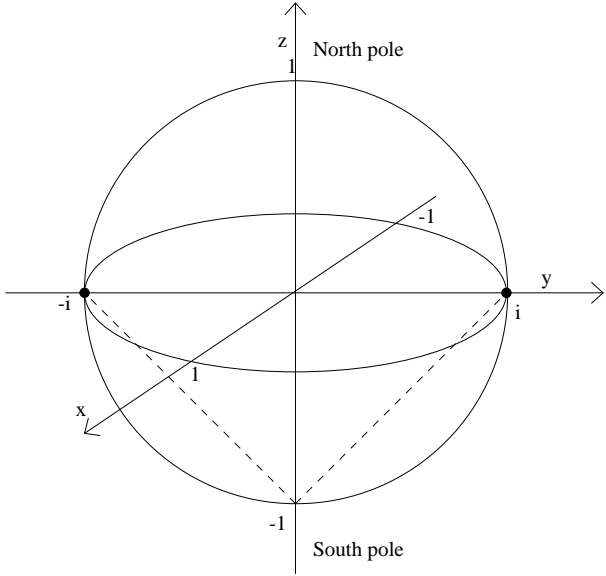}}}

\caption{\label{poincare2} Majorana representation of the state with 
polynomial $(z-i)(z+i)$}
\end{figure}

In the case where the degree $m$ of the polynomial is less than $2j$, $2j-m$ 
additional points at the South pole (the projective point) are added to the 
projection, corresponding to `roots at infinity'. Figure {\ref{poincare2}} 
depicts an example of stereographic projection for the spin-$1$ state 
represented by the quadratic polynomial 
\begin{equation}
p(z)=(z-i)(z+i)=z^2-1\leftrightarrow[|11\rangle+|1-1\rangle]/\sqrt{2}.
\end{equation}
By contrast, for spin 1, the maximum and minimum spin states $|1,1\rangle$ and
$|1,-1\rangle$ correpond to two repeated points at the North and South Pole,
respectively; on the other hand the pure intermediate spin state $|1,0\rangle
=|ud + du\rangle/\sqrt{2}$ corresponds to one point at the North pole and the 
other at the South pole. The Bell states, $|1,\pm\rangle\equiv|uu\pm dd\rangle
/\sqrt{2}$ are also antipodal points, but are located around the equator, 
specifically at coordinates (1,0,0)\& (-1,0,0) and (0,1,0) \& (0,-1,0). 
Under all accepted measures of entanglement, the spin zero state $|0,0\rangle
=|ud-du\rangle/\sqrt{2}$, the Bell states $|1,\pm \rangle$ and $|1,0\rangle$ 
(in fact, all local unitary transformations of these states) are maximally 
entangled. This suggests that taking the two points as ``far apart as one 
another'' on the Poincar\'{e} sphere is one possible way of producing maximal
entanglement. 

Another point of interest is that for these spin 1 states, the rotationally 
invariant dispersion measure,
$$(\Delta J)^2 \equiv \langle \vec{J}.\vec{J}\rangle - \langle \vec{J}\rangle.
                     \langle \vec{J}\rangle,$$
attains the maximum value of 2, because $\langle\vec{J}\rangle$ vanishes; so
it also suggests that the quantity $(\Delta J)^2/\langle \vec{J}\rangle^2$
might serve\cite{DeltaJ} as a way of characterising the entanglement of the
individual 1/2 spins that make up the $j$ state. In fact the dispersion in $J$
is equivalent to dispersion in any of the components $J^{(i)}$ and to the
covariance of any two components. This is because the total angular momentum
is $J=J^{(1)}+J^{(2)}+\dots+J^{(n)}$ and all states on the Poincar\'{e} sphere
are symmetrised; therefore the values taken by all the local $J^{(i)}$ are the
same. Thus when acting on these symmetrised states, $J=n J^{(i)}$, for all
$i=1,2,\ldots,n$, and the variance of any local $J^{(i)}$ is just as good an
entanglement measure. We can also take any two subspaces $i,j$ and evaluate
the covariance of $J^{(i)}$ and $J^{(j)}$; the space being symmetrised, any
$i,j$ (including $i=j$) may be taken! Thus the symmetrised states of maximised
$(\Delta J)^2$ have maximal covariance of the local $J^{(i)}$, as well. 
Finally, since we are dealing with spin 1/2 states, $J^{(i)}$ have only have
two distinct eigenvalues, and hence these local operators are directly 
proportional to the equal weight local operators of dimension two defined in 
section 3.

We may use these considerations to find the symmetrised states of maximum 
dispersion for arbitrarily high $J$ - these states will have the maximum 
covariance of any two local $J^{(i)}$. From the result in section 4, the local
covariance attains the maximum value of 1 whenever the reduced density 
matrices for the subspace on which those local operators jointly act are of 
maximum entropy. For a spin $\half$ space, the reduced density matrices must 
be of the form $\rho_i=\half[|u_i\rangle\langle u_i|+|d_i\rangle\langle d_i|]$.

For general $j$-values, observe that disentangled states on the Poincar\'{e} 
sphere are fully factorisable since the states are symmetrised. Clearly a 
fully factorisable state will have zero local operator variance for any 
operator with the local state as an eigenstate. Their Poincar\'{e} sphere 
representation simply consists of $n=2j$ repeated roots, since the general 
symmetrised, factorised state is a product of $n$ kets
$$(a|u\rangle+b|d\rangle))\ldots(a|u\rangle+b|d\rangle))\ldots
 \leftrightarrow (az + b)^n $$
because of the $z$-symmetrisation properties for spin $n$ systems.

Next it is useful to ask which states maximise $(\Delta J)^2$ if this is to
serve as a possible indication of entanglement. When $j=1$, as we have seen 
in section 3, the states are represented on the Poincar\'{e} sphere by two 
diametrically opposed points, one example being the state $|uu+dd\rangle/
\sqrt{2}$. When $j=3/2$ the states having maximum $(\Delta J)^2$ are those 
states which maximise the covariance of any two local operators. As was seen 
such states must be expressible as symmetrised unitary transformations of 
$(|uu\rangle+|dd\rangle)/\sqrt{2}$ in the local basis for the two operators 
in question. Thus the overall states must be expressible, by symmetry, as 
global rotations of $|uuu+ddd\rangle/\sqrt{2}$. These are the `triangular'
states, namely those represented by 3 points on the Poincar\'{e} sphere 
arranged in an equilateral triangle around any great circle---a global unitary
transformation (which preserves symmetrisation) simply rotates 
this configuration around the sphere. Choosing
the circle to lie equatorially, a polynomial producing such roots is
$p(z)=z^3+1$, which corresponds to the state $|\psi\rangle = [|3/2,3/2\rangle
+ |3/2,-3/2\rangle]/\sqrt{2} = |uuu+ddd\rangle/\sqrt{2}.$

\begin{figure}[tb]
\mbox{\centering \epsfxsize=8cm \epsfbox{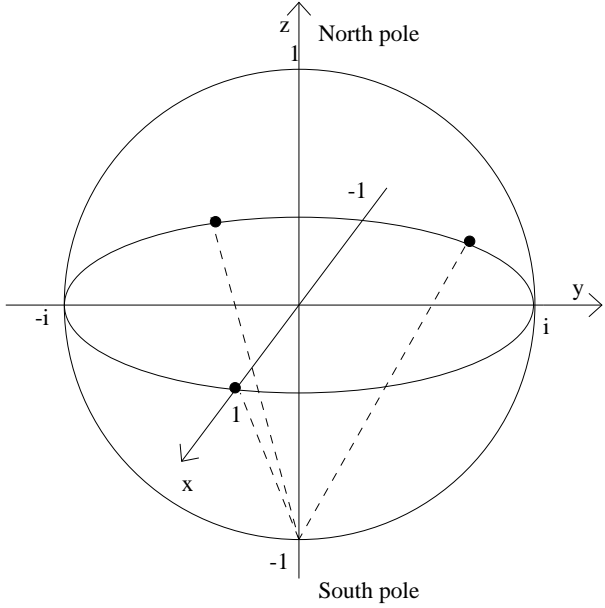}}
\caption{\label{poincare3} Majorana representation of the state with polynomial 
$p(z)=z^3+1$. }
\end{figure}

When $j=2$ the symmetrised states of maximum dispersion are the states with a tetrahedral
representation on the sphere. Choosing one of the apices of the tetrahedron
at the North Pole, we arrive at the polynomial
$$p(z)=z(z+\sqrt{2})(z-\sqrt{2}{\rm e}^{i\pi/3})(z-\sqrt{2}{\rm e}^{-i\pi/3})
  = z^4 +2\sqrt{2}z,$$
which corresponds to the maximally entangled state
$$|\psi\rangle = [|2,2\rangle +\sqrt{2}|2,-1\rangle]/3
  = |uuuu + uddd + dudd + ddud + dddu\rangle/\sqrt{5}. $$
Note that here, as in the previous case, the maximal $|\psi\rangle$ lead to
vanishing mean values $\langle\psi|\vec{J}|\psi\rangle.$

For $j=5/2$ there are two classes of states with maximum dispersion, with
slightly different geometries. The first of these is pyramidal with one
apex at the North pole and the other four apices at equal latitude (or any 
global rotation of this state), while the second has one point at the North 
pole another at the South Pole and the remaining three points distributed 
equilaterally on the equator. Hence the first configuration corresponds to 
the polynomial $p(z)= z^5 + z^4/\sqrt{3}$ or the state 
$|\psi\rangle= [|5/2,5/2\rangle+\sqrt{5/3}|5/2,-3/2\rangle]/\sqrt{8/3}$, 
whereas the second configuration leads to $p(z)=z^4+z$, or the state 
$[|5/2,5/2\rangle + |5/2,-3/2\rangle]/\sqrt{2}.$
Both choices have maximum $(\Delta J)^2$ and so the {\em total spin variance 
is unable to discriminate between them}. However, local equal-weight measures 
are able to discriminate between these states, so it seems that in more 
complicated cases entanglement is most naturally described by keeping the 
covariance of local operators in mind.

In order to use the (rotationally invariant) dispersion as a measure of pure 
state entanglement, one needs to consider that the symmetrised nature of the 
state is a reflection of the choice of basis. Thus one might define the 
variance-entanglement for a general state as the variance of the symmetrised 
form of the state, under an appropriate local unitary transformation. However,
it is not always possible to symmetrise an arbitrary state with local unitary 
transformations in spaces of spin 1 or higher. Thus it seems that the 
dispersion is not a perfect entanglement measure, although it does 
give an indication of the degree of entanglement of these particular states, 
because of its connection with the covariance through symmetrisation.

\section{An Integrity Basis for Density Matrix Invariants}
Consider $\rho$ for a composite $N^{(1)}N^{(2)}$-dimensional system as an
object transforming under $U(N^{(1)})\times U(N^{(2)})$ like 
$\{\bar{1}\}\{1\}\times \{\bar{1}\}\{1\}$, where $\{\bar{1}\}\{1\}$ denotes 
the reducible $N^2$ representation of $U(N)$; in tensor notation we can
write $\rho$ in the form $\rho_{ai}^{bj}$, where early Latin letters refer
to the first unitary group and the later letters stand for the second group.
Here we want to count the number of $U(N^{(1)})\times U(N^{(2)})$ singlets
$S_n$ in the symmetrised product $\rho^n$. Thus in the notation where
representations are labelled by partitions \cite{BKW},
$$\rho^n \sim (\{\bar{1}\}\{1\}\times \{\bar{1}\}\{1\})\otimes \{n\}
 \equiv \sum_{\kappa\circ\lambda\circ\mu\circ\nu\in \{n\}}
        \{\bar{\kappa}\}\{\lambda\}\times\{\bar{\mu}\}\{\nu\}.  $$
But $\{\bar{\kappa}\}\{\lambda\}$ can only contain a singlet if $\kappa\equiv
\lambda$. Moreover $\kappa=\lambda$ and $\mu=\nu$ have to be respectively
$N^{(1)}$ and $N^{(2)}$ part partitions, say ``$\kappa\vdash_{N^{(1)}} n$'',
etc., otherwise $\{\kappa\}$ vanishes in $U(N^{(1)})$. Therefore
$$\rho^n|_{\{0\}\times\{0\}}\subseteq \sum_{\kappa\vdash_{N^{(1)}}n,
  \lambda\vdash_{N^{(2)}}n} \{\bar{\kappa}\}\{\kappa\}\times
  \{\bar{\lambda}\}\{\lambda\}.$$
But it is known that $\alpha\circ\alpha\ni\{n\}$ for any $\alpha\vdash n$ and
moreover the order is immaterial, because the Clebsch series is symmetric.
Thus we have only to count the appropriate partitions,
$$S_n =\left|\rho^n\right|_{\{0\}\times\{0\}} = \left|\{\kappa,\lambda:
 \kappa\vdash_{N^{(1)}}n,\lambda\vdash_{N^{(2)}}n \}\right|.$$
This is not easy to work out in the general case, but is relatively simple for
the case $N^{(1)}=N^{(2)}=2$. When $n$ is even or odd, the partitions are:
$$n=2k:\qquad \{2k,0\}, \{2k-1,1\},\ldots \{k,k\}$$
$$n=2k+1:\qquad \{2k+1,0\}, \{2k,1\},\ldots, \{k+1,k\}.$$

Now the generating function for invariants of order $n$ in $\rho$ is written
$F(q):=\sum_{n=0}^\infty S_n q^n$. Including the even and odd cases,
$$F(q)=\sum_{k=0}^\infty [(k+1)^2q^{2k} + (k+1)^2 q^{2k+1}]
      = \frac{1+q^2}{(1-q^2)^2(1-q)}.$$
The denominator of $F(q)$ is crucial for its interpretation: we can
recognize that for the $2\times 2$ case invariants are freely generated by
two quadratic factors (namely $(1-q^2)^2$) and one linear factor (viz.
$(1-q)$), but there is an extra quadratic factor in the numerator which may
only be used once. We may associate these factors with
$${\rm Linear:}\qquad {\rm tr}\rho={\rm tr}_{(1)}{\rm tr}_{(2)}\rho=1\qquad
  {\rm anyway},$$
$${\rm Quadratic:}\qquad \chi_1\equiv{\rm tr}_{(1)}({\rm tr}_{(2)}\rho)^2 =
  \rho_{ai}^{bi}\rho_{bj}^{aj},$$
$$\qquad\qquad \&~\qquad \chi_2\equiv{\rm tr}_{(2)}({\rm tr}_{(1)}\rho)^2 =
  \rho_{ai}^{aj}\rho_{bj}^{bi},$$
$${\rm Extra~quadratic:}\qquad \epsilon_{aa'}\epsilon^{bb'}\epsilon_{ii'}
  \epsilon^{jj'} \rho_{bj}^{ai}\rho_{b'j'}^{a'i'}.$$
The last of these invariants is obviously related to $\chi_1, \chi_2$ and
tr($\rho^2$); however we do not get a new invariant from its square because
it is then expressible {\em entirely} as products of $\chi_1,\chi_2$,
tr $\rho^2$ and (tr $\rho)^2\equiv 1$. Thus effectively tr($\rho^2$) is 
{\em only allowed once}.

We conclude that {\em any} local unitary invariant entanglement measure in
the 2$\times$2 case must take the form:
$$E(\rho) := F(\chi_1,\chi_2) + {\rm tr}(\rho^2).G(\chi_1,\chi_2),$$
where $F, G$ are functions which depend on the way $E(\rho)$ is defined.
We think that this result must be useful for classifying entanglement.

\section*{Acknowledgement}
We thank V. Vedral for helpful discussions by electronic mail.

\end{document}